\documentclass[debug,overfull]{epl}

\title{Decrumpling membranes by quantum effects}
\author{M. E. S. Borelli\inst{1}\thanks{E-Mail:
\email{borelli@physik.fu-berlin.de}} \and
H. Kleinert\inst{1}\thanks{E-Mail: \email{kleinert@physik.fu-berlin.de, http://www.physik.fu-berlin.de/$\sim$kleinert}}}
\institute{
  \inst{1} \hspace{-0.2cm}Institut f\"ur
  Theoretische Physik, Freie Universit\"at Berlin, Arnimallee 14, 14195
  Berlin. 
}


\begin{document}

\maketitle

\begin{abstract}
  The phase diagram of an incompressible fluid membrane subject to
  quantum and thermal fluctuations is calculated exactly in a large
  number of dimensions of configuration space.  At zero temperature, a
  crumpling transition is found at a critical bending rigidity
  $1/\alpha_{\rm c}$.  For membranes of fixed lateral size, a
  crumpling transition occurs at nonzero temperatures in an auxiliary
  mean field approximation. As the lateral size $L$ of the membrane
  becomes large, the flat regime shrinks with $1/\ln L$.
\end{abstract}

\section{Introduction} \label{intro}

Amphiphilic molecules in aqueous solution form fluid bilayers with
vanishing surface tension.
This causes
them to undergo strong shape fluctuations, governed by the
Canham-Helfrich curvature energy \cite{canham,helf1}
\begin{equation}
{\cal H}_0 = \frac{1}{2\alpha_0} \int {\rm d}S \,H^2,
\label{helfmodel}
\end{equation}
where ${\rm d}S$ is the surface element, $H$ corresponds to 
the doubled mean curvature of the surface at each point, and $1/\alpha_0$ is
the bending rigidity.

Thermal undulations renormalize $1/\alpha_0$ as follows
\cite{helf3,peliti,klein1,forster,helf2,forster2,fluid}:
\begin{equation} \label{eq:alpha1}
\frac{1}{\alpha} = \frac{1}{\alpha_0} \left[ 1 - \frac{3}{4 \pi}
 k_{\rm B} T \alpha_0 \ln (\Lambda L) \right] ,
\end{equation}
where $\Lambda$ is an ultraviolet wavevector cutoff set by the inverse
width of the molecules in the membrane, and $L$ is an infrared cutoff
determined by its finite size.  In practice, membranes occur in
the form of spherical vesicles, and $L^2$ is determined by their surface
area.  At finite temperatures, the
model is only defined for finite planar surfaces.

For $L$ larger than the de Gennes-Taupin persistence length $\xi_{\rm
p} = \Lambda^{-1} \exp (4 \pi / 3 k_{\rm B} T \alpha_0)$ \cite{DGT},~the
renormalized  bending rigidity $1/\alpha$ vanishes. Beyond the persistence
length, the normal vectors of the surface are uncorrelated, and the membrane
is crumpled. The renormalization group flow extracted from the
perturbative result (\ref{eq:alpha1}) as well as nonperturbative
calculations in the large-$d$ limit \cite{kl,alonso,olesen,david}, where
$d$ is the dimension of the embedding space, exclude the possibility
of a phase transition in the Canham-Helfrich model, even for
tensionless membranes.

Recently, the model has been extended by a kinetic term to include
quantum fluctuations \cite{qmpert}.  A one-loop renormalization group
analysis showed that quantum effects stiffen the membrane.  The ground
state corresponds to a flat configuration, where the normal vectors of
the surface are strongly correlated.  The flat phase exists up to a
critical temperature
\begin{equation}
 \label{eq:tc_pert}
  T_{\rm c} = \frac{4 \pi}{3 k_{\rm B}} \frac{1}{\alpha}.
\end{equation}
Above $T_{\rm c}$, thermal fluctuations overcome quantum
effects, and the membrane is always crumpled.

In this paper, we analyze the behavior of the quantum membrane
exactly for very large dimension $d$ of the embedding space at all
temperatures.  Since the model is exactly solvable in this limit, we
can calculate all of its relevant properties, in particular its order
parameter and phase diagram.

\section{Definition of the model}

The surface describing the membrane is parametrized by a vector field
${\bf X}(\vec\sigma)$ in the $d$-dimensional embedding space, where
$\vec \sigma = (\sigma_1,\sigma_2)$ is a two-dimensional parameter
space.  In this parametrization, the Hamiltonian (\ref{helfmodel})
reads
\begin{equation}
{\cal H}_0 =\frac{1}{2 \alpha_0} \int {\rm d}^2 \sigma \sqrt{g}
(\Delta {\bf X})^2
\label{helfmodel2},
\end{equation}
where
\begin{equation} \label{gmunu}
g_{a b}= \partial_{a}{\bf X} \cdot \partial_{b}{\bf X}
\end{equation}
is the metric induced by the embedding, and $g \equiv \det[g_{a b}]$.
The symbol $\partial_a \, (a=1,2)$ denotes the derivative with respect
to the parameters $\sigma_1, \sigma_2$, and $\Delta =
g^{-1/2} \partial_a g^{a b} g^{1/2} \partial_b$ is the
Laplace-Beltrami operator.  As in Ref.\ \cite{qmpert}, we add to the
Hamiltonian (\ref{helfmodel2}) a kinetic term to account for quantum
fluctuations:
\begin{equation}
{\cal T} = \frac{1}{2 \nu_0} \int {\rm d}^2 \sigma \sqrt{g} \, \dot{\bf
X}^2,
\label{dynterm}
\end{equation}
where $\bf X$ is now time-dependent, $1/ \nu_0$ is the bare mass density, and
the dot indicates a time derivative.

The euclidean action describing the quantum membrane is thus
\begin{equation} \label{action}
S_0  =  \int {\rm d} \tau \, {\rm d}^2 \sigma \sqrt{g} \left[
\frac{1}{2\nu_0} \dot{\bf X}^2 + \frac{1}{2 \alpha_0}
(\Delta {\bf X})^2 \right],
\label{euclaction}
\end{equation}
and the partition function $Z$ can be represented as a functional integral over
all possible surface configurations ${\bf X}(\vec\sigma, \tau)$:
\begin{equation} \label{partfunc2}
Z = \int {\cal D}{\bf X} \exp(-S_0[{\bf X}]/\hbar).
\end{equation}

\section{Large-$\symbol{100}$ approximation} \label{larged}

For large $d$ it is useful to consider $g_{a b}$ as an
independent field \cite{polyakov}, and impose relation (\ref{gmunu})
with help of a Lagrange multiplier $\lambda_{a b}$.  We consider
the case where
the classical action will have an extremum around an almost flat
configuration.  In the $d$-dimensional generalization of the Monge
parametrization of an almost flat surface, the metric tensor becomes

\begin{equation}
\label{eq:gmunu_monge}
g_{a b}=\delta_{a b} + \partial_{a}{\bf X} \cdot
\partial_{b}{\bf X}.
\end{equation}
The partition function for the membrane can then be  written as
\begin{equation} \label{partfunc3}
Z = \int {\cal D}g \, {\cal D} \lambda \, {\cal D} {\bf X} \, {\rm
e}^{-S_0/\hbar},
\end{equation}
with the euclidean action
\begin{equation} \label{action2}
S_0 = \int {\rm d} \tau \, {\rm d}^2 \sigma \sqrt{g} \left\{ \frac{1}{2
\nu_0}\dot{\bf X}^2 + r_0 + \frac{1}{2 \alpha_0}\left[ (\Delta {\bf
X})^2 + \lambda^{a b}(\delta_{a b} + \partial_{a}{\bf X} \cdot
\partial_{b}{\bf X} - g_{a b}) \right] -
\frac{c_0}{4} \lambda_{aa}^2 \right\}.
\end{equation}

We have included a bare surface tension $r_0$ to absorb infinities
arising in the process of renormalization.  The renormalized, physical
surface tension $r$ will be set equal to zero at the end of our
calculations. We have further added a term proportional to
$\lambda_{aa}^2$, with the proportionality constant $c_0$ being the
in-plane compressibility of the membrane. This is also necessary to
absorb infinities,and the renormalized compressibility $c$ will be
set equal to zero at the end to describe an incompressible planar
fluid.

Note that the functional integration over the Lagrange multipliers
$\lambda_{ab}$ in (\ref{partfunc3}) has to be performed along the
imaginary axis for convergence.

The functional integral over all possible surface configurations $\bf
X(\vec\sigma, \tau)$ in (\ref{partfunc3}) is Gaussian, and can be immediately carried
out, yielding an effective action
\begin{equation} \label{effectaction}
S_{\rm eff} = \tilde{S}_0 + S_1,
\end{equation}
with
\begin{equation}
\tilde{S}_0 = \int {\rm d} \tau \, {\rm d}^2 \sigma \sqrt{g} \left[ r_0
+ \frac{\lambda^{a b}}{2 \alpha_0}(\delta_{a b} - g_{a b})  -
\frac{c_0}{4} \lambda_{aa}^2 \right]
\label{s0tilde},
\end{equation}
and
\begin{equation}
S_1 = \frac{\hbar}{2} d \, \mbox{Tr} \ln \left[- \partial^2_0
+ \frac{\nu_0}{\alpha_0}( \Delta^2 - \partial_a \lambda^{a b} \partial_b)
\right]. \label{s1}
\end{equation}

For large $d$, the partition function (\ref{partfunc3}) is dominated
by the saddle point of the effective action (\ref{effectaction}) with
respect to the metric $g_{a b}$ and the Lagrange multiplier
$\lambda^{a b}$, and we are left with a mean-field theory in these
fields.  For very large membranes, translational invariance allows us
to assume that this saddle point is symmetric and homogeneous
\cite{alonso,olesen,david,kleinld}, such that
\begin{equation}
g_{a b}= \varrho_0 \delta_{a b}; \; \lambda^{a b} = \lambda_0 g^{a b} =
\frac{\lambda_0}{\varrho_0} \delta^{a b},
\end{equation}
with constant $\varrho_0$ and $\lambda_0$. There, the functional trace in
(\ref{s1}) becomes an integral $\int {\rm d} \tau \, {\rm d}^2 \sigma$ $\int
{\rm d} \omega \, {\rm d}^2 q/(2 \pi)^3$ over the $(2+1)$-dimensional phase
space, after replacing $\partial_0^2 \to - \omega^2$ and ${g^{a b}
  \partial_{a} \partial_{b} \to - \bf q}^2$.

\section{Zero Temperature Properties} \label{zerotemp}

At zero temperature, the phase space integral in (\ref{s1})
yields
\begin{equation} \label{s1t0}
S_1 = \frac{\hbar}{2} d \int {\rm d}\tau {\rm d}^2 \sigma \varrho_0
  \sqrt{\frac{\nu_0}{\alpha_0}} \left\{  \frac{\Lambda^4}{8 \pi}  +
  \frac{\lambda_0}{8 \pi} \Lambda^2  + \frac{\lambda_0^2}{64 \pi} \left[ 1 - 2
  \ln \left( \frac{4 \Lambda^2}{\lambda_0} \right)\right]\right\},
\end{equation}
where the ultraviolet divergences of the integral have been
regularized by a wavevector cutoff $\Lambda$.  The first term in
(\ref{s1t0}) is a constant and renormalizes the surface tension to
\begin{equation}
  \label{eq:r0t0}
  r = r_0 + \frac{\hbar d}{16 \pi} \sqrt{\frac{\nu_0}{\alpha_0}} \Lambda^4.
\end{equation}
The quadratically divergent term renormalizes the bending rigidity in the
second term of (\ref{s0tilde}).  The logarithmically divergent term
proportional to $\lambda_0^2$ modifies the in-plane compressibility to
\begin{equation} \label{compressibility}
c = c_0 + \frac{\hbar d}{32 \pi} \sqrt{\frac{\nu_0}{\alpha_0}}
\ln \left( 4 e^{-1/2} \frac{ \Lambda^2}{\mu^2} \right),
\end{equation}
where $\mu$ is a renormalization scale.  We now set $r$ and $c$ equal
to zero, to describe a tensionless incompressible membrane.

The renormalized effective action is then 
\begin{equation}\label{seffld}
S_{\rm eff} = \int {\rm d}\tau \, {\rm d}^2 \sigma \varrho \lambda \left\{
  \frac{1}{\alpha} \left( \frac{1}{\varrho} -1 \right) +
   \frac{1}{\sqrt{\alpha \alpha_{\rm c}}} + \frac{a}{\sqrt{\alpha}}
  \lambda \left[ \ln \left( \frac{\lambda}{\bar{\lambda}}\right) - \frac{1}{2}\right]\right\},
\end{equation}
where we have defined the critical bending rigidity
\begin{equation}
  \label{alphact0}
  \frac{1}{\alpha_{\rm c}} \equiv \frac{\hbar^2 d^2 \nu_0}{256 \pi^2} \Lambda^4 .
\end{equation}
and the constants $a \equiv \hbar d \nu^{1/2}/(64 \pi)$, $\bar{\lambda}
\equiv \mu^2 e^{-1/2}$.  From the second-derivative matrix of $S_{\rm eff}$
with respect to $\varrho$ and $\lambda$ we find that the stability of
the saddle point is guaranteed only for $\lambda < \bar{\lambda}$.  Note
that  the integration over $\lambda_{ab}$ in (\ref{partfunc3})
along the imaginary axis requires
a maximum of (\ref{seffld}) with respect to $ \lambda $
for stability.

\begin{figure}
\begin{center}
\twofigures[width=5cm,height=4cm]{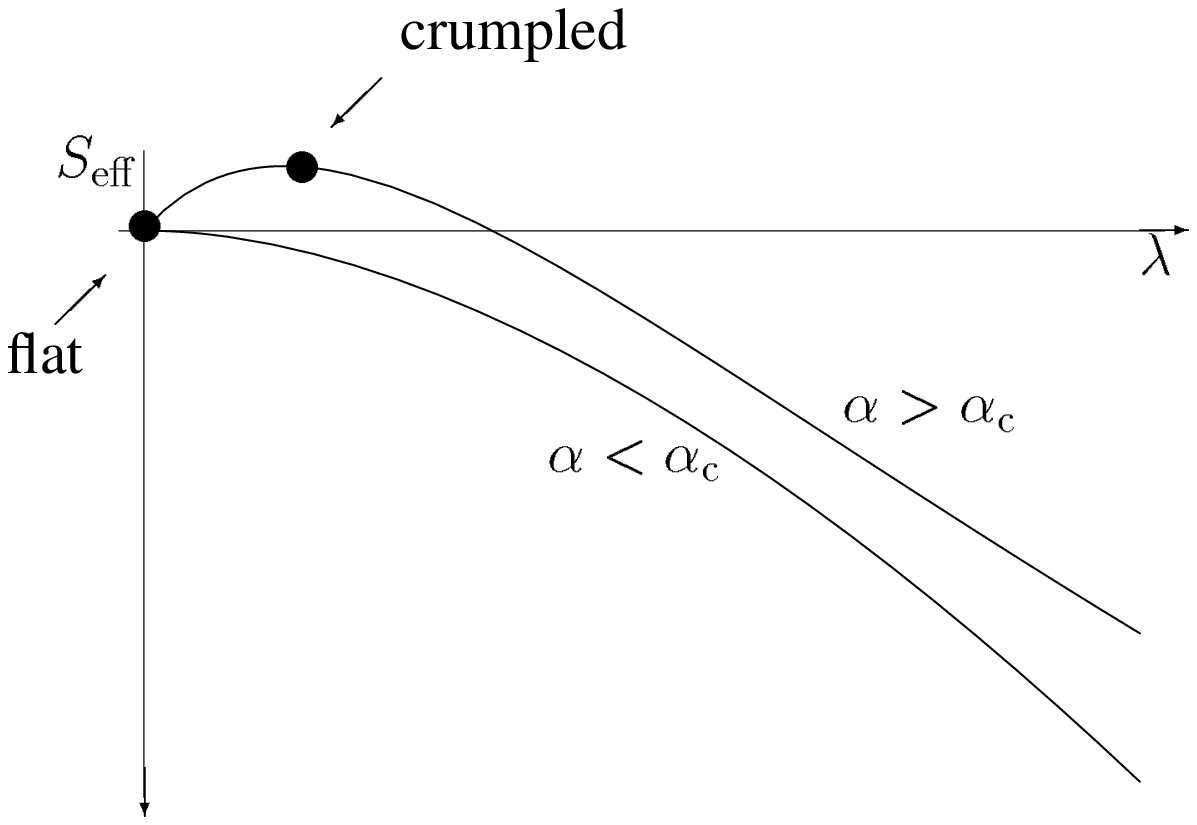}{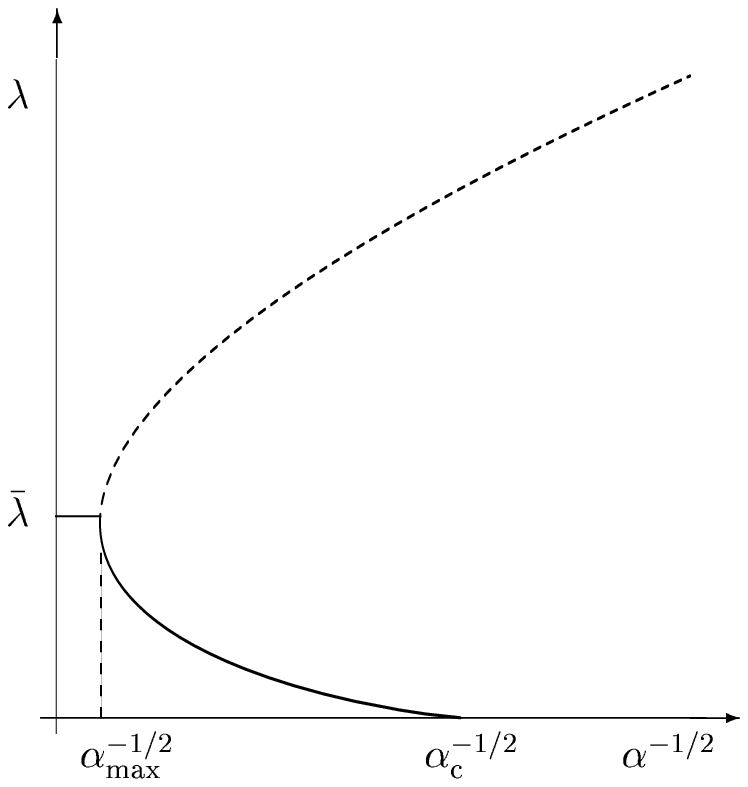}
\end{center}
\caption{Effective action at $T=0$ in units of $a$. The
physical solution of the saddle point equations (\ref{lambdasp1}),
(\ref{lambdasp2}) lies at the maxima.}
\label{fig:freezero}
\caption{Physical branch of the
solution of Eq.\ (\ref{lambdasp2}) for $\lambda$ as a function of the
stiffness $\alpha^{-1}$.  The dashed curve indicates the
unstable extremum of the action.}
\label{fig:lambda}
\end{figure}

The extremization of (\ref{seffld}) with respect to $\varrho$ leads to two
different solutions for the saddle point, namely
\begin{eqnarray}
  \label{lambdasp1}
  \lambda &=& 0 \,\,\,\,\, \mbox{or} \\
  \label{lambdasp2}
  \lambda \left[ \ln \left( \frac{\lambda}{\bar{\lambda}} \right) -
  \frac{1}{2} \right] &=& \frac{1}{a} \left( \frac{1}{\alpha^{1/2}} - \frac{1}{\alpha_{\rm
  c}^{1/2}} \right).
\end{eqnarray}
These describe two different phases existing at zero
temperature.  For $\alpha < \alpha_{\rm c}$, $\lambda = 0$ is the only
possible solution for the saddle point. This solution corresponds to
the flat phase, as we shall verify below. For $\alpha > \alpha_{\rm
c}$, on the other hand, there exists a solution of Eq.\
(\ref{lambdasp2}) for nonzero $\lambda$.  This solution corresponds to
the crumpled phase.  The behavior of the effective action
(\ref{seffld}) is shown in Fig.\ \ref{fig:freezero}.  As $\alpha$
approaches the critical point from below, i.\ e.\ the membrane
softens, $\lambda$ becomes nonzero, and the surface crumples. 

To determine the saddle point solution for $\varrho$ we extremize
(\ref{seffld}) with respect to $\lambda$. In the flat phase, where
$\lambda=0$, we obtain
\begin{equation}
  \label{rho1t0}
  \varrho_{-}^{-1} = 1 - \left( \frac{\alpha}{\alpha_{\rm c}}\right)^{1/2},
\end{equation}
showing that the total area of the membrane increases as $\alpha$
approaches $\alpha_{\rm c}$ from below, with a crumpling transition at
$\alpha_{\rm c}$. In the crumpled phase, $\varrho$ is given by
\begin{equation}
  \label{rho2t0}
  \varrho_{+}^{-1} = \left( \frac{\alpha}{\alpha_{\rm
  c}}\right)^{1/2}  - 1 - a \sqrt{\alpha} \lambda,
\end{equation}
with nonzero $\lambda$. As $\alpha$ approaches $\alpha_{\rm c}$ from
above, $\lambda$ tends to zero, and $\varrho$ goes again to infinity.  The
positivity of $\varrho$ and the stability of the saddle point imply that
there is an upper bound for the bending rigidity, given by
\begin{equation}
\frac{1}{\alpha_{\rm max}^{1/2}} = \frac{1}{\alpha_{\rm c}^{1/2}} - a
\bar{\lambda},
\end{equation}
below which an incompressible membrane becomes unstable.
 The behavior of
$\lambda$ in the two phases is shown in Fig.\
\ref{fig:lambda}. The behavior of $\varrho^{-1}$ is shown
 in Fig.\ \ref{fig:rho}.

\section{Finite-Temperature Properties} \label{finitetemp}

At finite temperature, the phase space integral in (\ref{s1}) involves
a sum over the Matsubara frequencies \cite{kleinbook}:
\begin{equation} \label{fun:matb}
\omega_n = 2 \pi n k_{\rm B}T/\hbar , \;\;\;\;\; n = 0, \pm 1, \pm 2 \cdots.
\end{equation}
For small $\lambda_0$, a series expansion leads to

\begin{eqnarray}
  \label{s1ft}
  S_1 = \int {\rm d}\tau \, {\rm d}^2 \sigma \varrho_0 \Bigg\{
  \!\!\!\!\!\! &&
    \frac{\hbar d}{16 \pi} \sqrt{\frac{\nu_0}{\alpha_0}} \Lambda^4  -
  \frac{\hbar d \pi}{24} \sqrt{\frac{\alpha_0}{\nu_0}} \left(
\frac{k_{\rm B}  T}{\hbar}\right)^2 + \frac{\hbar d}{16 \pi} \lambda_0
    \sqrt{\frac{\nu_0}{\alpha_0}} \Lambda^2 \nonumber \\ &&
+ \lambda_0 \frac{d k_{\rm B} T}{8 \pi} \ln
  \left( L^2 \frac{k_{\rm B} T}{\hbar} \sqrt{\frac{\alpha_0}{\nu_0}}\right)
    + a \frac{\lambda_0^2}{\sqrt{\alpha_0}} \left[3 - 2 \gamma + 2
  \ln \left( \frac{\lambda_0}{8 \pi} \frac{L^2}{\Lambda^2} \frac{k_{\rm B}
  T}{\hbar} \sqrt{\frac{\alpha_0}{\nu_0}} \right) \right] \nonumber \\ &&
 + \frac{\hbar d}{2}\sqrt{\pi} \sum_{m=3}^{\infty}
  \frac{(-1)^{m+1}\lambda_0^m}{m \, 2^{2 m} \pi^m} \left(\frac{\hbar}{k_{\rm B}
      T}\right)^{m-2} \left( \frac{\nu_0}{\alpha_0}\right)^{\frac{m-1}{2}}
  \frac{\Gamma(\frac{m-1}{2})}{\Gamma(\frac{m}{2})} \zeta(m-1) \Bigg\}
\end{eqnarray}
As in the zero-temperature discussion, we absorb the logarithmic
divergence by renormalizing the in plane compressibility via Eq.\
(\ref{compressibility}), setting $c$ equal to zero for incompressible
membranes.  The surface tension receives now a temperature dependent
renormalization
\begin{equation}
  r = r_0 + \frac{\hbar d}{16 \pi} \sqrt{\frac{\nu_0}{\alpha_0}} \Lambda^4 -
 \frac{\hbar d \pi}{24} \sqrt{\frac{\alpha_0}{\nu_0}} \left(
\frac{k_{\rm B}  T}{\hbar}\right)^2,
\end{equation}
and $r_0$ is chosen to make $r=0$ for tensionless membranes at all temperatures.

Extremization of the renormalized combined effective action
 (\ref{s0tilde}) and (\ref{s1ft}) with respect to $\varrho$ leads
 again to two possible solutions for the saddle point, namely $
 \lambda = 0$ or $\lambda = \lambda_{\rm T}$, with
\begin{eqnarray}
  \label{lambdaT}
&& \lambda_{\rm T} \left[ \ln \left( \frac{\lambda_{\rm T}}{\bar{\lambda}} \right) -
  \frac{1}{2} \right] + \lambda_{\rm T} \left[ 1 - \gamma + \ln
  \left(\frac{L^2}{8 \pi} \frac{k_{\rm B} T}{\hbar} \sqrt{\frac{\alpha}{\nu}}
  \right)\right]
  \nonumber \\ &&+ 32 \pi^{3/2} \sum_{m=3}^{\infty} \frac{ (-1)^{m+1} \lambda_{\rm
  T}^{m-1}}{m \, 2^{2 m} \pi^m} \left(\frac{\hbar}{k_{\rm B}
  T}\right)^{m-2} \left( \frac{\nu}{\alpha}\right)^{\frac{m-2}{2}}
  \frac{\Gamma(\frac{m-1}{2})}{\Gamma(\frac{m}{2})} \zeta(m-1) \nonumber \\
  &&= \frac{1}{a} \left[ 1 - \alpha^{1/2} \left( 1 - \sqrt{\frac{\alpha_{\rm
  c}}{\alpha_{\rm  T}}} \right)\right] \left( \frac{1}{\alpha^{1/2}} -
  \frac{1}{\alpha_{\rm  T}^{1/2}}\right)
\end{eqnarray}
where
\begin{equation}
  \label{alphacT} \frac{1}{{\alpha_{\rm T}}^{1/2}} \approx
  \frac{1}{ \alpha_{\rm c}^{1/2}}\left[ \frac{1}{2} + \sqrt{\frac{1}{4} +
  \frac{d k_{\rm B} T }{8 \pi} \alpha_{\rm c}\ln
  \left(L^2\frac{k_{\rm B} T}{\hbar} \sqrt{\frac{\alpha_{\rm c}}{\nu}}
  \right)} \, \right]
\end{equation}
is the critical bending rigidity at finite temperature.
Alternatively, we find the critical temperature at a fixed bending
rigidity $1/\alpha$:
\begin{equation}
  \label{eq:tc_ld}
  T_{\rm c} \ln
  \left( L^2 \frac{k_{\rm B} T_{\rm c}}{\hbar} \sqrt{\frac{\alpha}{\nu}}
  \right) = \frac{8 \pi}{d k_{\rm B}}\left( \frac{1}{\alpha} -
  \frac{1}{\sqrt{\alpha \alpha_{\rm c}}} \right),
\end{equation}
in qualitative agreement with the perturbative critical temperature in
Eq.\ (\ref{eq:tc_pert}).

For $\alpha < \alpha_{\rm  T}$, Eq.\ (\ref{lambdaT})
has no solution for $\lambda_{\rm T}$. In this case, the membrane is
in the flat phase, the only available solution for the saddle
point being $\lambda = 0$.  For $\alpha > \alpha_{\rm T}$,
$\lambda_{\rm T}$ is nonzero, and the membrane is crumpled.

Let us now examine the saddle point solutions for $\varrho$. In the
crumpled phase where $\lambda = \lambda_{\rm T}$ is nonzero, we may
expand the effective action into a high-temperature series.
Extremization with respect to $\lambda_T$ leads to
\begin{eqnarray} \label{rhoplus}
\varrho_{+}^{-1} &=& \left[\left(\frac{\alpha}{\alpha_{\rm  T}} \right)^{1/2} -1
  \right]  \left[ 1 - \alpha^{1/2} \left( 1 - \sqrt{\frac{\alpha_{\rm
  c}}{\alpha_{\rm  T}}} \right)\right] - a \sqrt{\alpha} \lambda_{\rm T}
\nonumber \\
&-& \hbar d \alpha \frac{\sqrt{\pi}}{2}
\sum_{m=3}^{\infty} \frac{(-1)^{m+1}\lambda_{\rm T}^{m-1}}{2^{2 m} \pi^m} \left(
1 - \frac{2}{m} \right) \left(\frac{\hbar}{k_{\rm B} T}\right)^{m-2}
\left( \frac{\nu}{\alpha}\right)^{\frac{m-1}{2}}
\frac{\Gamma(\frac{m-1}{2})}{\Gamma(\frac{m}{2})} \zeta(m-1) .
\end{eqnarray}
The positivity of $\varrho$ and the stability of the
saddle point again define an upper bound for the inverse bending rigidity,
given by
\begin{equation}
\frac{1}{\sqrt{\alpha_{\rm max}^{T}}} =  \frac{1}{\sqrt{\alpha_{\rm max}}}
    \left( \frac{1}{2} + \frac{1}{2} \sqrt{1 - \frac{T}{T_{\rm stab}}} \right)
\end{equation}
with
\begin{equation}
k_{\rm B} T_{\rm stab} = \frac{4 \pi}{d \alpha_{\rm max} \ln(16
\pi/\hbar d \sqrt{\nu \alpha_{\rm c}} \bar{\lambda})}.
\end{equation}
For temperatures lower than $T_{\rm stab}$ the effective action
becomes unstable if the rigidity is lower than $1/\alpha_{\rm max}^{\rm
T}$. Above $T_{\rm stab}$, the membrane is stable at all
rigidities.

In the flat phase, the situation is more delicate.  For
$\lambda = 0$, $\varrho$ can be calculated exactly,
and we obtain
\begin{equation} \label{rhoftl0}
\varrho_{-}^{-1} = 1 - \alpha \frac{d k_{\rm B} T}{8 \pi} \ln \left[
\frac{\sinh \left( \frac{16 \pi}{d k_{\rm B} T} \frac{1}{\sqrt{\alpha \alpha_{\rm c}}} \right)}{\frac{\hbar}{ 2 k_{\rm B} T} \sqrt{\frac{\nu}{\alpha}} L^{-2}}\right],
\end{equation}
with an infrared regulator $L$ equal to the inverse lateral size of the
membrane.  For low temperatures, (\ref{rhoftl0}) may be approximated
by
\begin{equation}
\label{rhominus}
\varrho_{-}^{-1} \approx  \left[1 - \left(\frac{\alpha}{\alpha_{\rm  T}} \right)^{1/2}  \right]  \left[ 1 - \alpha^{1/2} \left( 1 - \sqrt{\frac{\alpha_{\rm
  c}}{\alpha_{\rm  T}}} \right)\right].
\end{equation}

At high temperatures, however, the positivity of $\varrho$ is not
guaranteed. For fixed, but high temperatures, and for fixed membrane
lateral size $L$, there is a characteristic value of the inverse
bending rigidity defined by
\begin{equation}
\label{Tcr}
\alpha^* = \frac{8 \pi}{d k_{\rm B} T \ln(16
\pi L^2 /\hbar d \sqrt{\nu \alpha_{\rm c}}) },
\end{equation}
above which $\varrho$ changes sign, and (\ref{rhoftl0}) is no longer
applicable.  Interestingly, for all $L$ and at all temperatures $T$, the
critical bending rigidity $1/\alpha_{\rm T}$ is larger than $1/\alpha^*$, so
that the crumpling transition still occurs.  The behavior of $\varrho$ is
depicted in Fig.\ \ref{fig:rho} below.
\begin{figure}
\begin{center}
\twofigures[width=6cm,height=5cm]{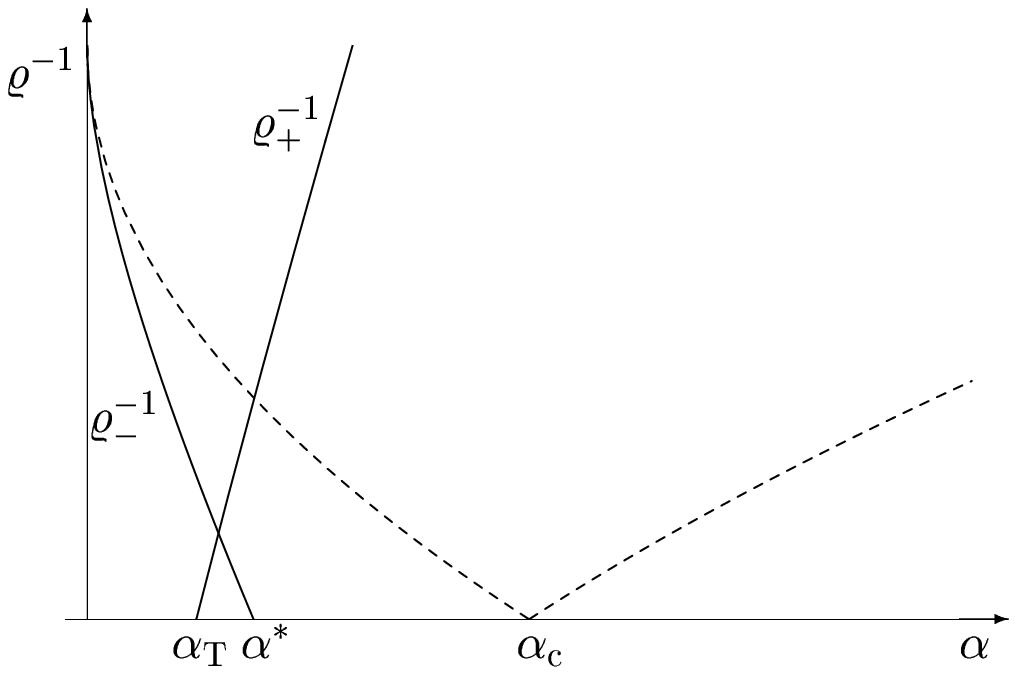}{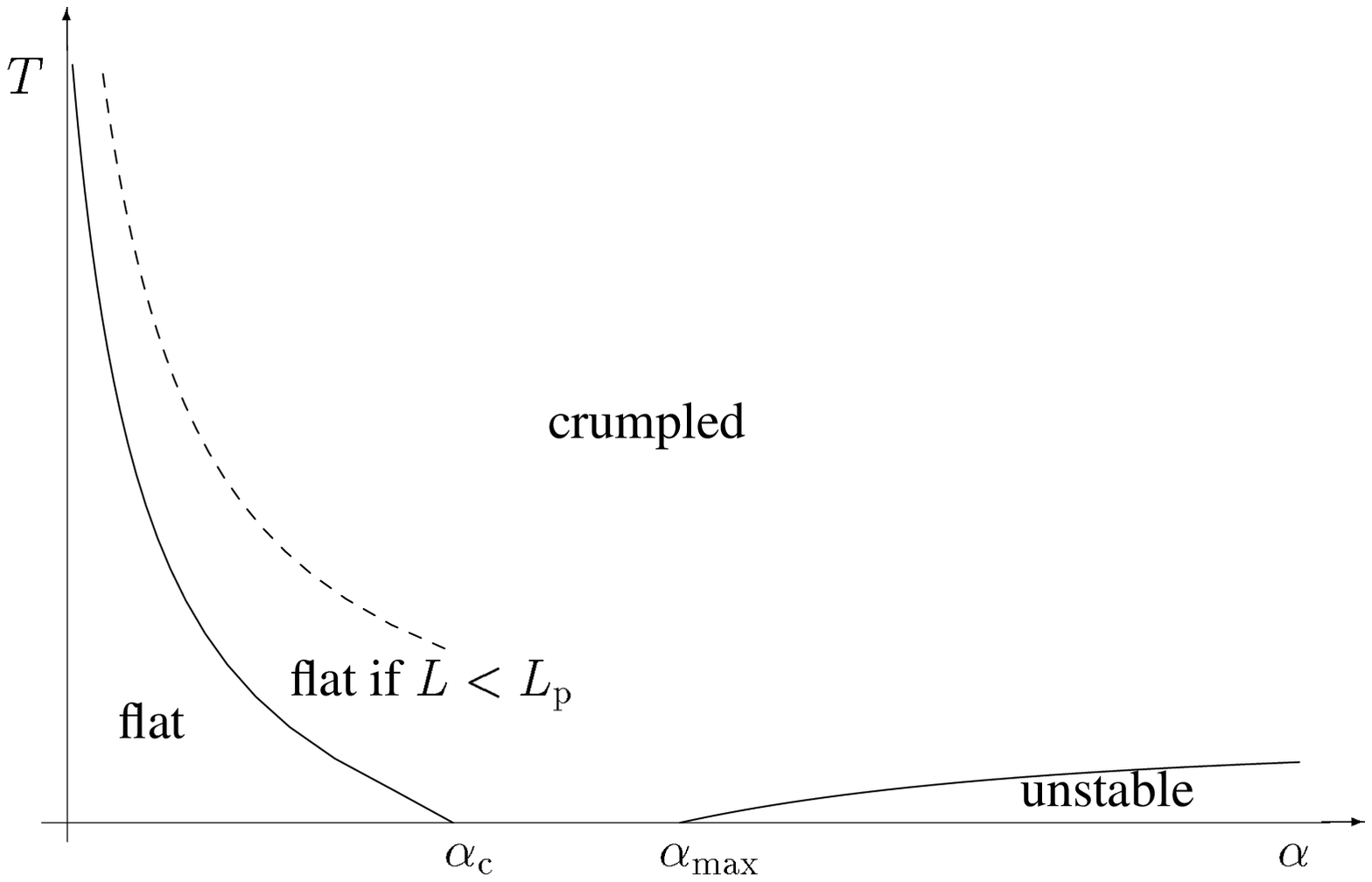}
\end{center}
\caption{ Behavior of $\varrho^{-1}$ for fixed $L^2$. In the flat
phase, $\varrho$ is given by (\ref{rhoftl0}), and in the crumpled
phase by (\ref{rhoplus}). The transition happens at $\alpha_{\rm
T}$, where the flat phase becomes crumpled. The dashed lines show the
behavior of $\varrho^{-1}$ at zero temperature.}
\label{fig:rho}
\caption{Phase diagram of the quantum membrane.  At $T=0$, there is a
crumpling transition as the rigidity $1/ \alpha $ falls below $1/
\alpha _c$.  For membranes of fixed lateral size $L$, the crumpling
transition still takes place at higher temperatures.  The critical
inverse rigidity tends asymptotically to zero as the temperature goes
to infinity. Below the dotted line the membrane is still flat if its
lateral size is smaller than the persistence length.  The unstable
region disappears for nonzero in-plane compressibility.}
\label{fig:phase}
\end{figure}

Note that Eq.\ (\ref{Tcr}) reflects the existence of a persistence length. At
fixed temperature, for $\alpha_{\rm T} < \alpha < \alpha^*$, the
membrane is flat at scales smaller than
\begin{equation}
L_{\rm p} = \Lambda^{-1} \exp\left( \frac{4 \pi}{d k_{\rm B} T \alpha }\right),
\end{equation}
and  crumpled at larger scales.  This agrees with the de Gennes-Taupin
persistence length $\xi_{\rm p}$.

As the projected area $L^2$ of the membrane approaches infinity, the root
$\alpha_{\rm T}$ of the branch $\varrho_{+}^{-1}$ (see Fig.\ \ref{fig:rho})
goes to zero, and the branch $\varrho_{-}^{-1}$ becomes unphysical.

The phase diagram of the quantum membrane is plotted in Fig.\
\ref{fig:phase}.

As the lateral size $L$ of the membrane goes to infinity, the inverse critical
bending rigidity $\alpha_{\rm T}$ goes to zero, and the crumpling transition
is washed out.  In the limit of infinite area, the ratio $\alpha^*/\alpha_{\rm
  T} = 1$.  The membrane is crumpled at large scales, and flat at scales
smaller than the persistence length.  Its behavior can thus be described by
the classical Canham-Helfrich model alone.

Let us finally characterize the two phases in terms of the correlation
functions between the normal vectors to the surface of the
membrane. For the solution $\lambda = 0$, this correlation function
coincides with the one we found perturbatively for the zero
temperature case, namely
\begin{equation}
\langle \partial_a {\bf X}(\vec \sigma,\tau) \cdot \partial_b {\bf X}(\vec \sigma',\tau)
\rangle \sim \frac{\delta_{a b}}{|\vec \sigma - \vec \sigma'|^3}.
\end{equation}
This solution corresponds to the flat, low temperature phase, where
the normal vectors are strongly correlated.  Such behavior can
incidentally also be obtained in the large-$d$ limit at high
temperatures by adding curvature terms of higher orders to the
Canham-Helfrich Hamiltonian (\ref{helfmodel}) to stabilize a negative
bending rigidity $1/\alpha_0$ \cite{carlo1,carlo2,carlo3}.

For nonzero $\lambda = \lambda_{\rm
T}$ the correlation function behaves as
\begin{equation}
\langle \partial_a {\bf X}(\vec \sigma,\tau) \cdot \partial_b {\bf X}(\vec \sigma',\tau)
\rangle \sim \delta_{a b}\exp(- \sqrt{\lambda_{\rm T}}|\vec \sigma - \vec \sigma'|).
\end{equation}
In this case, the normals to the membranes are uncorrelated beyond
a length scale $\lambda_{\rm T}^{-1/2}$. The exponential decay of the
correlation function shows that this solution corresponds to the
crumpled phase. The length scale $\lambda_{\rm T}^{-1/2}$
may also be identified as the persistence length $\xi_{\rm p}$
\cite{david,leibler}.

\section{Summary}

We have analyzed the temperature behavior of a membrane subject to
thermal and quantum fluctuations in the limit of large embedding space
dimension. We found that at zero temperature there is a crumpling
transition at some critical stiffness $1/ \alpha_{\rm c}$.  For
membranes of finite lateral size, quantum fluctuations are still
relevant and a crumpling transition occurs also at nonzero temperature
in an auxiliary mean field approximation As the lateral size of the
membrane goes to infinity, the transition disappears, and the membrane
is always crumpled in spite of quantum fluctuations, this being a
consequence of the infrared divergences in the flat phase.

\end{document}